\documentclass{aastex631}

\begin{document}

\title{The First Photometric Study of the Binary System CSS J003106.8+313347}

\author[0000-0001-9746-2284]{Ehsan Paki}
\affiliation{Binary Systems of South and North (BSN) Project, Tehran, Iran}

\author[0009-0004-8426-4114]{Sabrina Baudart}
\affiliation{Société Astronomique de France, Paris, France}

\author[0000-0002-0196-9732]{Atila Poro}
\affiliation{Astronomy Department of the Raderon AI Lab., BC., Burnaby, Canada}

\begin{abstract}
We performed the first photometric study of the CSS J003106.8+313347 W Ursae Majoris (W UMa)-type system based on ground-based observations. We extracted times of minima from our observations and proposed a linear ephemeris based on the increasing incline of the orbital period using a Markov chain Monte Carlo (MCMC) approach. The PHOEBE Python code and the MCMC approach were used for the light curve analysis. This system did not need starspots for the light curve analysis. Mass ratio, fillout factor, and inclination were obtained as $0.699$, $0.322$, and $60.6^{\circ}$ respectively. We also estimated the absolute parameters of the system using the Gaia DR3 parallax method. Therefore, the masses, radii, and luminosities have been determined to be $M_1=1.675$, $M_2=1.171$, $R_1=1.292$, $R_2=1.097$, $L_1=1.348$, and $L_2=1.221$. The orbital angular momentum ($J_0$) of the CSS J003106.8+313347 illustrates that this system is located in a region of contact binaries. The positions of the primary and secondary components on the Hertzsprung-Russell (HR) diagram are depicted.
\end{abstract}


\vspace{1.5cm}
\section{Introduction}
W UMa-type binary systems (EWs) variable stars are short orbital period eclipsing binaries in which both the primary and secondary components have filled their Roche lobes. Therefore, these systems are exchanging mass and energy with each other. W UMa-type systems play a substantial role in studying the physical characteristics of stars and their evolution (\citealt{1994ASPC...56..228B}, \citealt{2003MNRAS.342.1260Q},
\citealt{2005ApJ...629.1055Y},
\citealt{2007ApJ...662..596L}, \citealt{2008MNRAS.386.1756E},  \citealt{2012JASS...29..145E}).

The CSS J003106.8+313347 eclipsing binary system has an apparent magnitude of $V=14.73$ and is located in the northern hemisphere with coordinates R.A. $00^{\circ}$ $31’$ $06.8257”$ (J2000) and Dec. $+31^{\circ}$ $33’$ $47.1867”$ (J2000). Both the magnitudes and coordinates are from the Simbad\footnote{\url{http://simbad.u-strasbg.fr/simbad/}} database. The system is inserted as an EW type in the VSX\footnote{\url{https://www.aavso.org/vsx/}} database with an orbital period of 0.343666 days.

We have presented a new ephemeris, analyzed the light curve, and also estimated the absolute parameters of the system based on the Gaia DR3 parallax method. The paper is arranged in the following sections: In Section 2, the details of photometric observations and the data reduction method are given. Section 3 presents the minima extracted in this study and the new ephemeris of the CSS J003106.8+313347 system. The photometric light curve solutions and estimation of the absolute parameters of the system are propounded in Section 4. Finally, Section 5 includes the conclusion.

%%%%%%%%%%%%%%%%%%%%%%%%%%%%%%%%%%%%%%%%%%%%%%%%%%
\vspace{2.5cm}
\section{Observation and Data Reduction}
We have done the observation in a private observatory located in Toulon, France, at a longitude of $05^{\circ}$ $54’$ $35”$ E and a latitude of $43^{\circ}$ $8’$ $59”$ N, and an altitude of 68 meters above mean sea level. The system was observed on two nights on September 30, 2022, and November 5, 2022, with an apochromatic refractor TS optics with a 102mm aperture (520mm focal length), and a ZWO ASI 1600MM CCD with a $V$ standard filter. A total of 293 images were taken, each of which was $1\times1$ binned with a 110-second exposure time. The average CCD temperature during observations was $-15^{\circ}C$.

Gaia DR3 2859621459407078912 star at coordinates R.A. $00^{\circ}$ $32’$ $8.56”$ (J2000) and Dec. $+31^{\circ}$ $42’$ $54.1$ (J2000) with a magnitude of $14.613$, and Gaia DR3 2859715467651208320 star at coordinates R.A. $00^{\circ}$ $32’$ $19.43”$ (J2000) and Dec. $+31^{\circ}$ $45’$ $40.3$ (J2000) with a magnitude of $14.391$ have been considered comparison and check stars, respectively. Both the magnitudes and coordinates are from the Simbad database.

The basic data reduction was carried out for the bias, dark, and flat fields of each CCD image with Muniwin (v2.1.32). Finally, we normalized all data using the AstroImageJ software (\citealt{2017AJ....153...77C}).

%%%%%%%%%%%%%%%%%%%%%%%%%%%%%%%%%%%%%%%%%%%%%%%%%%
\vspace{1.5cm}
\section{New Ephemeris}
\begin{figure*}
\begin{center}
\includegraphics[scale=0.42]{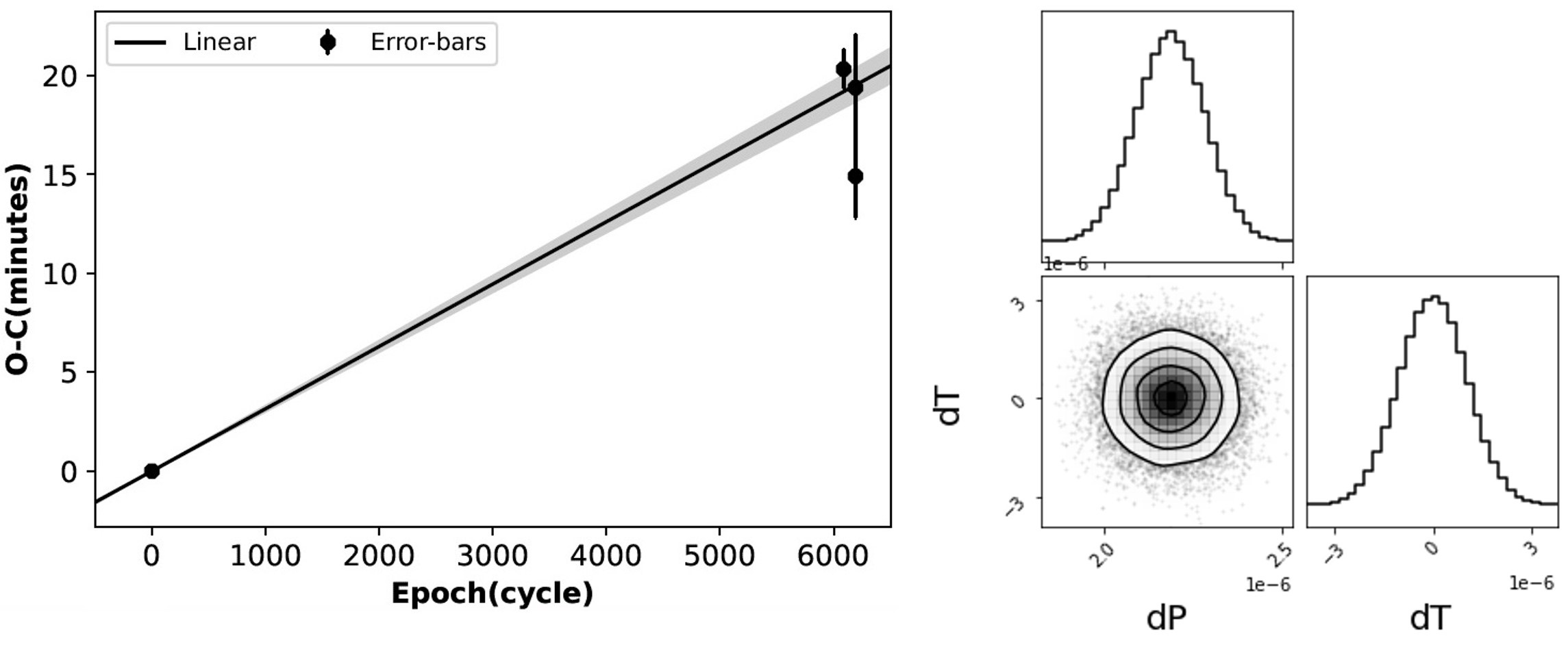}
    \caption{Left: The O-C diagram of the eclipsing binary with a linear (black line) model. The shaded regions show the model parameters’ 68th percentile values, while the curve represents their median values. Right: the MCMC-obtained corner plot.}
\label{Fig1}
\end{center}
\end{figure*}

We extracted one primary and two secondary minima, from our observations with a Gaussian function using a Python code (Table \ref{tab1}). Also, there is just one mid-eclipse time for CSS J003106.8+313347 in the literature that we could find. All times of minima have been specified in Barycentric Julian Date in Barycentric Dynamical Time ($BJD_{TDB}$). We determined the variations between the Observed mid-eclipse times from their Calculated values (O-C). The mid-eclipse timings ($t_0$) and orbital periods ($P$) used for this were taken from the ASAS-SN catalog (\citealt{jayasinghe2019asas}) and were 2457761.72387 ($BJD_{TDB}$) and $0.3436651^d$, respectively. The O-C diagram of the system is shown in Figure \ref{Fig1}. Due to the small number of minima, only a linear fit was carried out on O-C values. Finally, we used the PyMC3 package in Python to calculate a new ephemeris based on the MCMC method (\citealt{2016ascl.soft10016S}). The MCMC sampling was performed using 20 walkers, 10000 iterations for each walker, and a 1000 burn-in period. The new ephemeris obtained from Equation \ref{eq1}:

\begin{equation}
\label{eq1}
BJD_{TDB}(Min.I)=2457761.72387(1)+0.3436673(1)\times E
\end{equation}

 where $E$ is the integer number of orbital cycles after the reference mid-eclipse time.

\begin{table}
\caption{Available times of minima for CSS J003106.8+313347.}
\centering
\begin{center}
\footnotesize
\begin{tabular}{c c c c c}
 \hline
 \hline
 Min.($BJD_{TDB}$) & Error & Epoch & O-C & Reference\\
\hline
2457761.72387 & & 0 & 0 & \cite{jayasinghe2019asas}\\
2459853.45562 & 0.00068	& 6086.5 & 0.01412 & This study\\
2459889.36797 & 0.00187 & 6191 & 0.01346 & This study\\
2459889.53669 & 0.00146 & 6191.5 & 0.01036 & This study\\
\hline
\hline
\end{tabular}
\end{center}
\label{tab1}
\end{table}

%%%%%%%%%%%%%%%%%%%%%%%%%%%%%%%%%%%%%%%%%%%%%%%%%%
\vspace{1.5cm}
\section{Light Curve Analysis}
\begin{figure*}
\begin{center}
\includegraphics[scale=0.64]{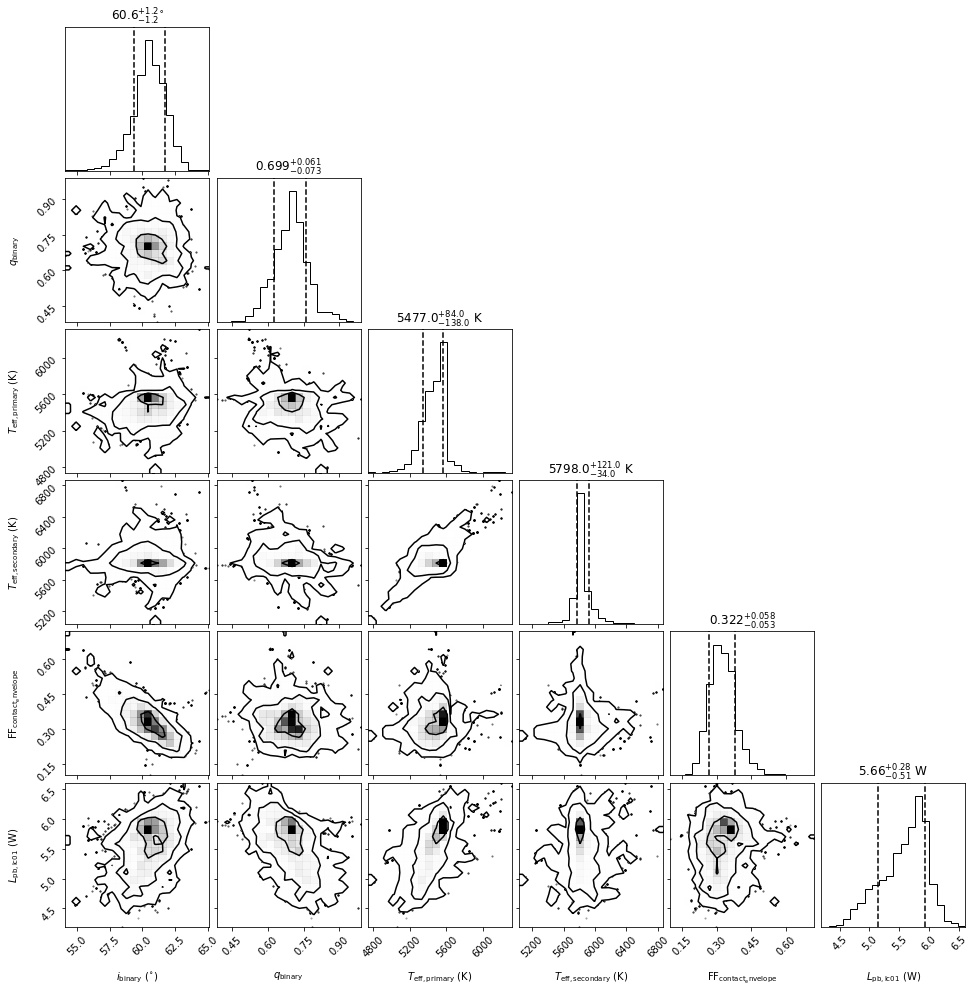}
    \caption{The corner plots of the system were determined by MCMC modeling.}
\label{Fig2}
\end{center}
\end{figure*}

\begin{figure*}
\begin{center}
\includegraphics[scale=0.64]{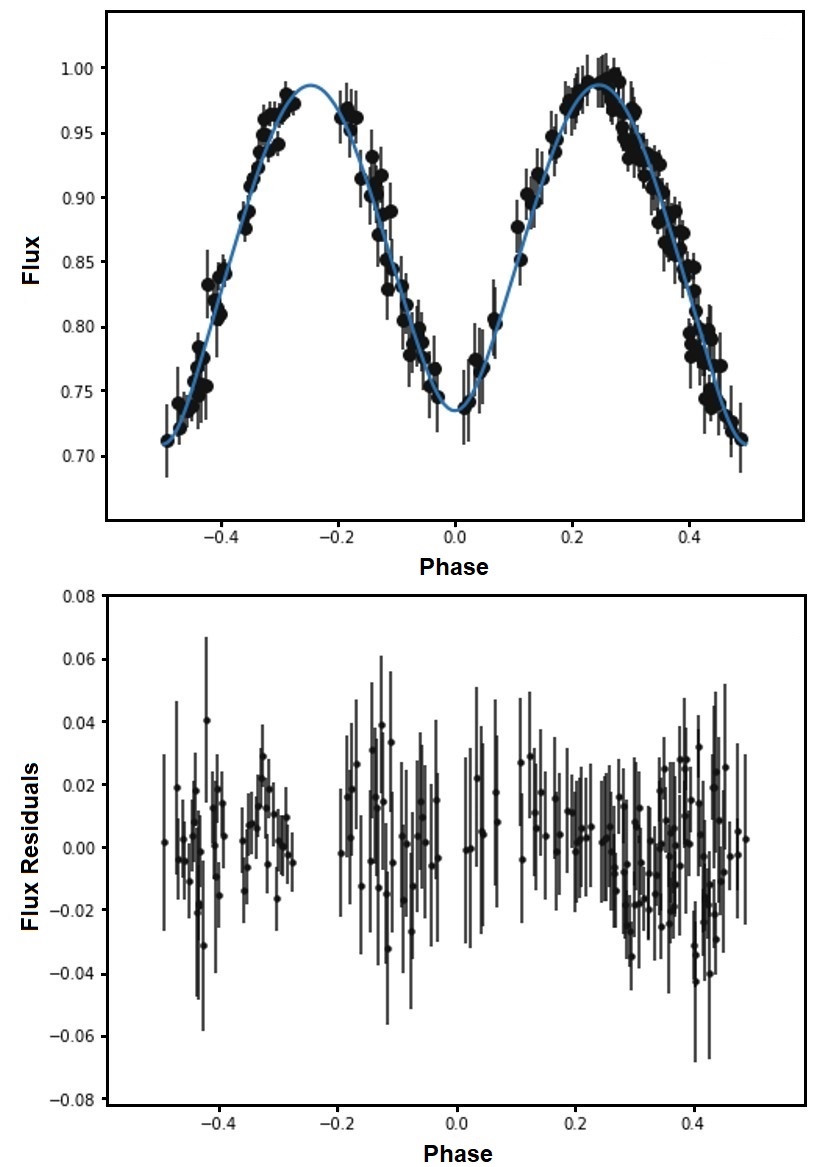}
    \caption{The observed light curve of the system (black dots); and the synthetic light curve was obtained from the light curve solution in the $V$ filter. The orbital phase was retained but the relative flux shifted arbitrarily.}
\label{Fig3}
\end{center}
\end{figure*}

\begin{figure*}
\begin{center}
\includegraphics[scale=0.38]{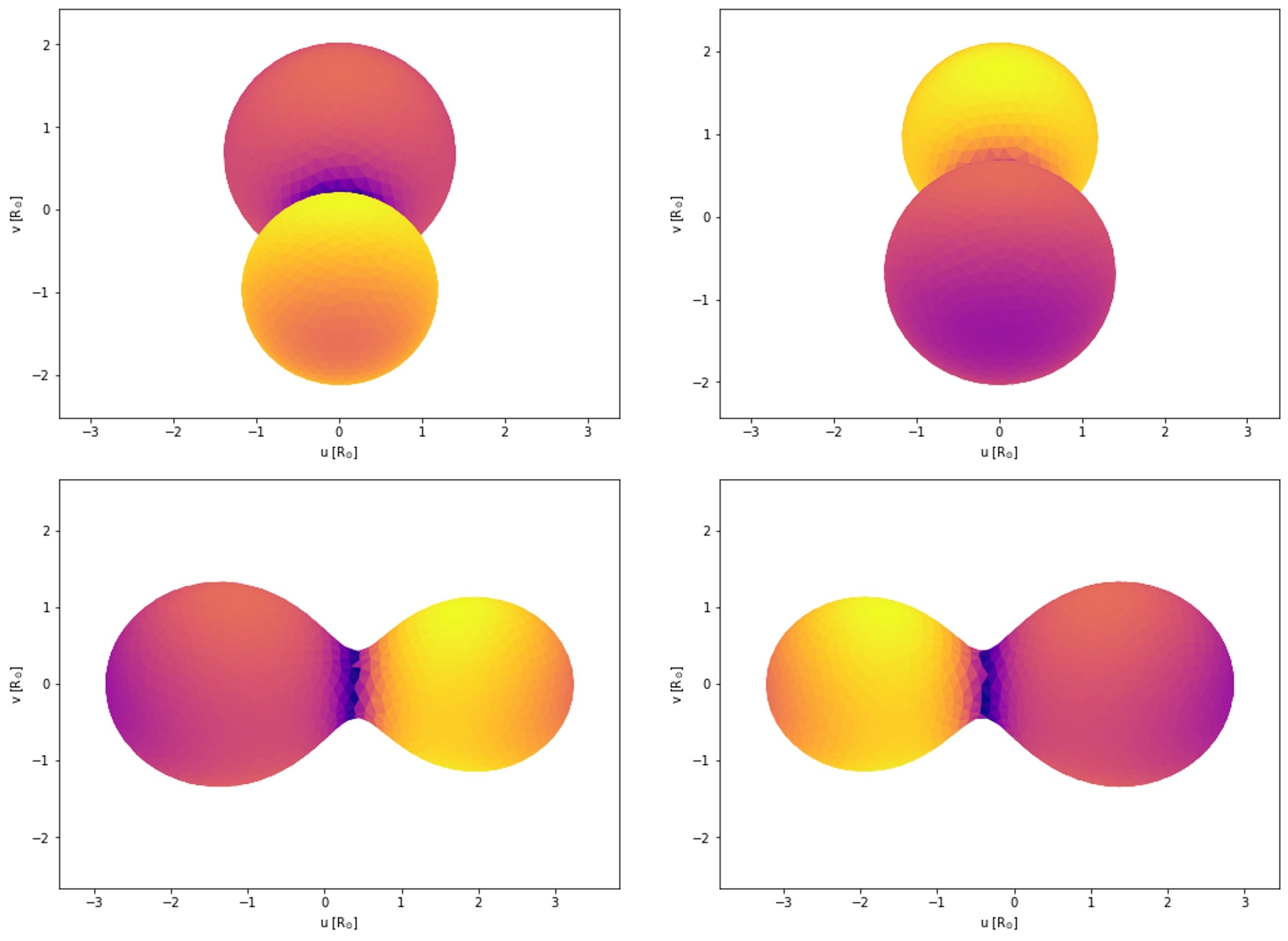}
    \caption{3D representations of the CSS J003106.8+313347 system based on the light curve solution.}
\label{Fig4}
\end{center}
\end{figure*}

Light curve analysis of the CSS J003106.8+313347 system has been carried out with the PHOEBE 2.4.7 version and the MCMC approach (\citealt{2020ApJS..250...34C}).
The Gaia DR3 temperature was considered as the effective temperature of the secondary component according to the morphology of the light curve. The gravity-darkening coefficients and the bolometric albedo were assumed to be $g_1=g_2=0.32$ (\citealt{1967ZA.....65...89L}) and $A_1=A_2=0.5$ (\citealt{1969AcA....19..245R}), respectively. The \cite{2004A&A...419..725C} method was used to model the stellar atmosphere and the limb darkening coefficients were adopted as a free parameter in the PHOEBE. Since radial velocity studies have not been performed on this system, the main challenge was to estimate the mass ratio of the system. So, we performed $q$-search for different values of inclination ($i$), and fillout factor ($f$). The acquired value has been used as the initial value of the parameters in the MCMC process. Due to the symmetry of the maxima in the light curve, there was no need to apply starspots for the light curve solution. After applying a proper visual fitting to the light curve with the initial parameters, we used PHOEBE’s optimization tool to improve the theoretical fit. Finally, considering a normal Gaussian distribution in the range of solutions for inclination, mass ratio, fillout factor, effective temperature for primary and secondary components, and the luminosity of primary, we extracted the final values of the parameters along with their uncertainties using MCMC method based on the emcee package in PHOEBE (\citealt{2018ApJS..236...11H}). The light curve solution results are listed in Table \ref{tab2}. We did MCMC processing with 48 walkers and 500 iterations for each walker. The corner plots and final synthetic light curve was shown in Figure \ref{Fig2} and Figure \ref{Fig3} respectively. Figure \ref{Fig4} displays the component positions for four different phases of an orbital period.
In Figure \ref{Fig4}, PHOEBE sets the colors based on the star's surface temperature; therefore, the brighter the color, the higher the temperature; also the connection point of the component stars has a lower temperature due to the gravity darkening (\citealt{2016ApJS..227...29P}).

\begin{table}
\caption{Photometric solution of CSS J003106.8+313347.}
\centering
\begin{center}
\footnotesize
\begin{tabular}{c c c}
 \hline
 \hline
Parameter && Result\\
\hline
$T_{1}$ (K) && $5477_{\rm-(138)}^{+(84)}$\\
\\
$T_{2}$ (K) && $5798_{\rm-(34)}^{+(121)}$\\
\\
$q=M_2/M_1$ && $0.699_{\rm-(73)}^{+(61)}$\\
\\
$\Omega_1=\Omega_2$ && $3.11(13)$\\
\\
$i^{\circ}$ &&	$60.6_{\rm-(1.2)}^{+(1.2)}$\\
\\
$f$ && $0.322_{\rm-(53)}^{+(58)}$\\
\\
$l_1/l_{tot}$ && $0.509_{\rm-(24)}^{+(12)}$\\
\\
$l_2/l_{tot}$ && $0.491_{\rm-(24)}^{+(12)}$\\
\\
$r_{1(mean)}$ && $0.439(12)$\\
\\
$r_{2(mean)}$ && $0.377(14)$\\
\\
Phase shift && $0.044(1)$\\
\hline
\hline
\end{tabular}
\end{center}
\label{tab2}
\end{table}

%%%%%%%%%%%%%%%%%%%%%%%%%%%%%%%%%%%%%%%%%%%%%%%%%%
\vspace{1.5cm}
\section{Absolute Parameters}
The Gaia DR3 parallax method was used for estimating absolute parameters (\citealt{2022MNRAS.510.5315P}a). We utilized $V_{max}=14.88\pm0.1$ from the ASAS-SN catalog, the extinction coefficient $A_v=0.155$ from the \cite{2011ApJ...737..103S} study and the system’s distance $d(pc)=1512\pm66$ from Gaia DR3. The absolute magnitude $M_v$ of the system and subsequently the absolute magnitude for the primary and secondary components was calculated by Equations \ref{eq2} and \ref{eq3}.

\begin{equation}\label{eq2}
M_v=V-5log(d)+5-A_v
\end{equation}

\begin{equation}\label{eq3}
M_{v(1,2)}-M_{v(tot)}=-2.5log(\frac{l_{(1,2)}}{l_{(tot)}})
\end{equation}

The bolometric magnitude $M_{bol}$ of each component of the binary was obtained by Equation \ref{eq4}.

\begin{equation}\label{eq4}
M_{bol}=M_{v}+BC
\end{equation}

where the bolometric correction for the primary and secondary components $BC_1=-0.144\pm0.027$ and $BC_2=-0.077\pm0.014$ were derived as a function of the effective temperature of the stars (\citealt{1996ApJ...469..355F}). The luminosity and radius of primary and secondary, $L_{1,2}$, $R_{1,2}$, and separation between the center of mass of the components were calculated by the following Equations (\ref{eq5}, \ref{eq6} and \ref{eq7}):

\begin{equation}\label{eq5}
M_{bol}-M_{bol_{\odot}}=-2.5log(\frac{L}{L_{\odot}})
\end{equation}

\begin{equation}\label{eq6}
R=(\frac{L}{4\pi \sigma T^{4}})^{1/2}
\end{equation}

\begin{equation}\label{eq7}
a=\frac{R}{r_{mean}}
\end{equation}

In addition, having mass ratio from the results of light curve analysis, the mass of each component was obtained by the well-known Kepler’s third law (Equation \ref{eq8}). The surface gravity was determined by Equation \ref{eq9}. All characterized absolute parameters were given in Table \ref{tab3}. It should be noted that the average is used if the absolute parameters' uncertainties have different upper and lower limits.

\begin{equation}\label{eq8}
\frac{a^3}{G(M_1+M_2)}=\frac{P^2}{4\pi^2}
\end{equation}

\begin{equation}\label{eq9}
g=G_{\odot}(M/R^2)
\end{equation}

\begin{table}
\caption{The absolute parameters of the CSS J003106.8+313347 system in the order of estimates of parameters.}
\centering
\begin{center}
\footnotesize
\begin{tabular}{c c c c c}
 \hline
 \hline
Parameter & & Primary & & Secondary\\
\hline
$M_v(mag.)$ && 4.560(43) && 4.600(45)\\
$M_{bol}(mag.)$ && 4.416(70) && 4.523(59)\\
$L(L_\odot)$ && 1.348(88) && 1.221(66)\\
$R(R_\odot)$ && 1.292(95) && 1.097(60)\\
$a(R_\odot)$ && 2.926(93) &&\\
$M(M_\odot)$ && 1.675(94) && 1.171(176)\\
$log(g)(cgs)$ && 4.440(88) && 4.426(112)\\
\hline
\hline
\end{tabular}
\end{center}
\label{tab3}
\end{table}

%%%%%%%%%%%%%%%%%%%%%%%%%%%%%%%%%%%%%%%%%%%%%%%%%%
\vspace{1.5cm}
\section{Conclusion}
\begin{figure*}
\begin{center}
\includegraphics[scale=0.54]{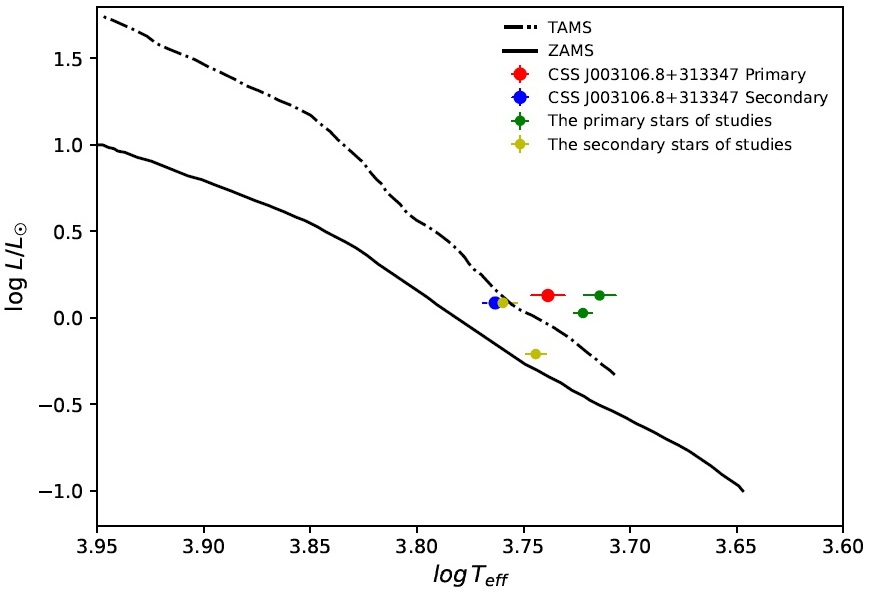}
    \caption{The positions of the primary and secondary components on the HR diagram in which the theoretical ZAMS and TAMS curves are indicated for the CSS J003106.8+313347 system as well as two systems from literature.}
\label{Fig5}
\end{center}
\end{figure*}

The CSS J003106.8+313347 system was observed for two nights at an observatory in southern France. The minima were extracted from the observational light curves. The epoch and O-C values were calculated based on the reference ephemeris, and a linear fitting was applied. Therefore, we presented a new ephemeris, and it can be seen that there is an incremental linear fit in the O-C diagram.
The light curve analysis of the system was performed using the PHOEBE Python code. Based on the observational parameters and the light curve solution, the absolute parameters of the system were estimated.
According to the light curve solution, the secondary star is hotter than the primary star by 321 K. Stars’ temperatures represent that the primary and secondary spectral types are G8 and G5, respectively (\citealt{2018MNRAS.479.5491E}). The position of each of the primary and secondary components, as well as two systems from the literature, AH Mic from the \cite{poro2022first}b study and NR Cam from the \cite{tavakkoli2015first} study, has been determined in the HR diagram (Figure \ref{Fig5}). The HR diagram shows that the primary component is above the Terminal-Age Main Sequence (TAMS) and the secondary star is between the Zero-Age Main Sequence (ZAMS) and TAMS which is consistent with the binary stars’ evolution theories.

The orbital angular momentum of the system is $52.08\pm0.04$. This result is based on the equation presented by \cite{2006MNRAS.373.1483E} as follows:

\begin{equation}\label{eq10}
J_0=\frac{q}{(1+q)^2} \sqrt[3] {\frac{G^2}{2\pi}M^5P}
\end{equation}

\begin{figure*}
\begin{center}
\includegraphics[scale=0.54]{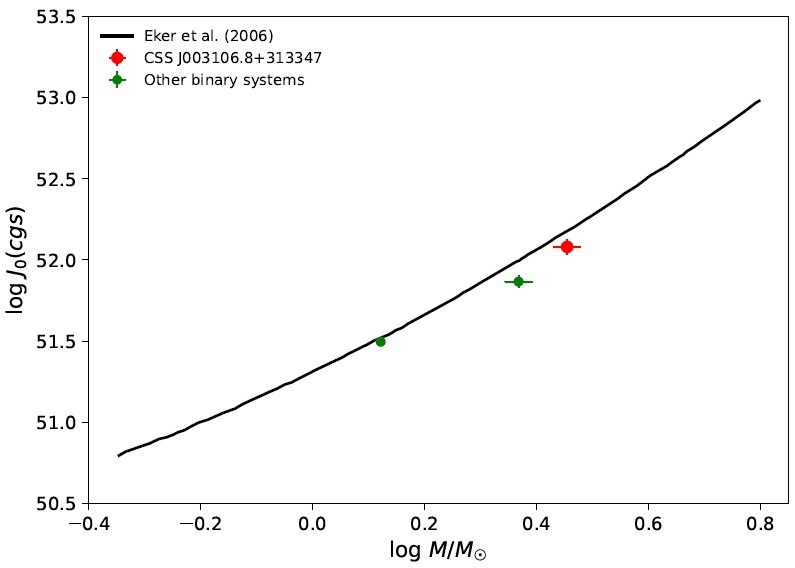}
    \caption{The positions of the CSS J003106.8+313347 on the $logJ_0-logM$ diagram as well as two contact systems from literature.}
\label{Fig6}
\end{center}
\end{figure*}

Where $q$ is the mass ratio, $M$ is the total mass of the system, $P$ is the orbital period, and $G$ is the gravitational constant.
The $logJ_0-logM$ diagram illustrates the orbital angular momentum of the CSS J003106.8+313347 in a contact binary systems region (Figure \ref{Fig6}).
\\
\\
According to the light curve solution and the estimation of the absolute parameters of the CSS J003106.8+313347, it can be concluded that this system is an overcontact binary.

%%%%%%%%%%%%%%%%%%%%%%%%%%%%%%%%%%%%%%%%%%%%%%%%%%
\vspace{1.5cm}
\section*{Acknowledgements}
This manuscript was prepared by the Binary Systems of South and North (BSN) project (\url{https://bsnp.info/}). We have made use of Gaia DR3 results. The Gaia mission is from the European Space Agency (ESA) (\url{http://cosmos.esa.int/gaia}), processed by the Gaia Data Processing and Analysis Consortium (DPAC).

%%%%%%%%%%%%%%%%%%%%%%%%%%%%%%%%%%%%%%%%%%%%%%%%%%
\vspace{1.5cm}
\section*{ORCID iDs}
\noindent Ehsan Paki: 0000-0001-9746-2284\\
Sabrina Baudart: 0009-0004-8426-4114\\
Atila Poro: 0000-0002-0196-9732\\

%%%%%%%%%%%%%%%%%%%%%%%%%%%%%%%%%%%%%%%%%%%%%%%%%%
\clearpage
%%%%%%%%%%%%%%%%%%%%%%%%%%%%%%%%%%%%%%%%%%%%%%%%%%
\vspace{1.5cm}
\bibliography{References}{}
\bibliographystyle{aasjournal}

\end{document}